# Superconductivity in the Ru-Doped CuIr$_2$Te$_4$ Telluride Chalcogenide

Dong Yan[a], Lingyong Zeng[a], Yishi Lin[b], Junjie Yin[c], Yuan He[a], Xing Zhang[a], Meiling Huang[a], Meng Wang[c], YihuaWang[b], Daoin Yao[c], Huixia Luo[a]*

[a]*School of Material Science and Engineering and Key Lab Polymer Composite & Functional Materials, Sun Yat-Sen University, No. 135, Xingang Xi Road, Guangzhou, 510275, P. R. China*

[b]*Department of Physics, Fudan University, Shanghai, 200433, China*

[c]*School of Physics, State Key Laboratory of Optoelectronic Materials and Technologies, Sun Yat-Sen University, No. 135, Xingang Xi Road, Guangzhou, 510275, P. R. China*

*Corresponding author/authors complete details (Telephone; E-mail:) (+0086)-2039386124*

*luohx7@mail.sysu.edu.cn*

**Abstract**

Here we report the effect of structural and superconductivity properties on Ru doped $CuIr_2Te_4$ telluride chalcogenide. XRD results suggest that the $CuIr_{2-x}Ru_xTe_4$ maintain the disordered trigonal structure with space group $P\bar{3}m1$ (No. 164) for $x \leq 0.3$. The lattice constants, $a$ and $c$, both decrease with increasing Ru content. Temperature-dependent resistivity, magnetic susceptibility and specific-heat measurements are performed to characterize the superconducting properties systematically. Our results suggest that the optimal doping level for superconductivity in $CuIr_{2-x}Ru_xTe_4$ is $x = 0.05$, where $T_c$ is 2.79 K with the Sommerfeld constant γ of 11.52 mJ mol$^{-1}$ K$^{-2}$ and the specific-heat anomaly at the superconducting transition, $\Delta C/\gamma T_c$, is approximately 1.51, which is higher than the BCS value of 1.43, indicating $CuIr_{1.95}Ru_{0.05}Te_4$ is a strongly electron-phonon coupled superconductor. The values of lower $\{H_{c1}(0)\}$ and upper $\{H_{c2}(0)\}$ critical field calculated from isothermal magnetization $\{M(H)\}$ and magneto-transport $\{\rho(T, H)\}$ measurements are 0.98 KOe and 2.47 KOe respectively, signifying that the compound is clearly a type-II superconductor. Finally, a "dome-like" shape superconducting $T_c$s vs. $x$ content phase diagram is established, where the charge density wave disappears at $x = 0.03$ while superconducting transition temperature ($T_c$) rises until it reaches its peak at $x = 0.05$, then, with decreasing when $x$ reaches 0.3. This feature of the competition between CDW and the superconductivity could be caused by tuning the Fermi surface and density of states with Ru chemical doping.

**Keywords**: $CuIr_{2-x}Ru_xTe_4$; Superconductivity; Charge density wave; Quaternary telluride chalcogenide.

**Introduction**

The group of $AB_2X_4$ materials, with metallic $A$ and $B$ elements and $X$ a chalcogen (O, S, Se, Te), has attracted much attention since it offers a versatile range of relevant physical properties. Generally speaking, the oxyspinels ($AB_2O_4$) are semiconductors with antiferromagnetic interactions, whereas the sulphospinels exhibit a much richer physical properties, such as metallic conduction, ferromagnetic, superconductivity, semiconductivity as well as antiferromagnetic interactions and so on.[1-8] Especially, the copper chalcogenide ($CuB_2X_4$) spinels have attracted remarkable attention due to their peculiar superconductivity and magnetism.

Copper chalcogenide $CuIr_2S_4$, for example, exhibits a temperature-induced metal-insulator (M-I) transition at 226 K, which is highly possibly attributed to the dimerization between Ir ions and the Jahn-Teller effect.[9-13] However, the isostructural $CuIr_2Se_4$ spinel remains metallic at ambient pressure, while above 4 GPa it exhibits semi-conductive behavior in the temperature range of 7-300 K.[9, 14] On the other hand, $CuRh_2S_4$ and $CuRh_2Se_4$ spinel are well known as superconductors with $T_c$ = 4.35 K and $T_c$ = 3.50 K, respectively.[15-19] Strikingly, copper chalcogenide spinel $CuV_2S_4$ superconducts at 4.45-3.20 K and shows three charge density wave (CDW) states ($T_{CDW1}$ = 55 K, $T_{CDW2}$ = 75 K, $T_{CDW3}$ = 90 K).[20-21]

It is well known that chemical doping can efficiently tune the crystal and electronic structures of copper chalcogenide spinels, leading to the formation of novel physical properties. For example, the M-I transition was decreased with the increase of Se substitution for S at $X$-site of $CuIr_2S_4$ or Rh substitution for Ir at $B$-site of $CuIr_2S_4$.[24-25] Besides, on Zn substitution for Cu in the $Cu_{1-x}Zn_xIr_2S_4$ solid solution, the M-I transition can be suppressed and superconductivity appears, with a maximum $T_c$ of 3.4 K near $x$ = 0.3.[27] Moreover, the superconductivity can be observed for $Cu(Ir_{1-x}Pt_x)_2Se_4$ ($0.1 \leq x \leq 0.35$) with a maximum $T_c$ = 1.76 K near $x$ = 0.2 with Pt substitution for Ir in the $CuIr_2Se_4$ solid solution.[28]

Unlike $CuB_2X_4$ sulpho- or seleno-compounds with cubic spinel structure, $CuB_2X_4$ copper chalcogenide telluro-compounds tend to possess lower dimensional structure. Recently, some reports suggested that low dimensionality leads to special electronic structures and allows relatively strong fluctuations, which may enhance superconductivity, even though charge-density wave (CDW) sometime competes, especially in the quasi-one-dimensional case.[22] Intrigued by this issue, we recently have systematically studied the properties of $CuIr_2Te_4$, which adopts a disordered trigonal structure with space group $P\bar{3}m1$,[23] and found coexistence of the superconducting ($T_c$ = 2.5 K) and CDW ($T_{CDW}$ = 250 K) in the copper telluride chalcogenide $CuIr_2Te_4$.[24] According to our previous calculation study, we find both orbital projected band structure and density of state, the bands near the Fermi energy $E_F$ mainly come from Te $p$ and Ir $d$ orbitals, similar to that of $CuIr_2S_4$ in spinel structure.[24] Therefore, it is reasonable to tune superconductivity properties by tuning the Fermi energy $E_F$ of $CuIr_2Te_4$ telluride chalcogenide with chemical doping. In this article we report the synthesis and physical properties of the $B$-site substitution solid solution $CuIr_{2-x}Ru_xTe_4$ ($0.0 \leq x \leq 0.3$). The structural properties of the $AB_2X_4$-type telluro-compounds $CuIr_{2-x}Ru_xTe_4$ ($0.0 \leq x \leq 0.3$) was characterized via X-ray diffraction (XRD). We characterize the effect of Ru substitution on the superconducting transition through temperature-dependent resistivity, magnetic susceptibility and specific-heat measurements. All measurements consistently confirm that the optimal doping level for superconductivity in $CuIr_{2-x}Ru_xTe_4$ is $x$ = 0.05. The specific-heat anomaly at the superconducting transition, $\Delta C/\gamma T_c$, is approximately 1.51, which indicates that $CuIr_{1.95}Ru_{0.05}Te_4$ is a strong electron-phonon coupling BCS type superconductor. A "dome-like" shape electronic phase diagram between charge density wave (CDW) and superconducting transition temperature $T_c$ versus Ru doing content $x$ has been established experimentally for this system. The CDW was immediately suppressed even with small amount Ru doping at $x$ = 0.03 while the superconducting transition temperature ($T_c$) rises until it reaches its peak at $x$ = 0.05, then, with decreasing when $x$ reaches 0.3. With discovery of this doping superconductor of $CuIr_2Te_4$, we found the effective method to improve the $T_c$ and also provides guidance for us to study other doping systems. This feature of the competition between CDW and the superconductivity could be induced by tuning the Fermi surface and density of states with Ru chemical doping.

**Experimental Section**

Polycrystalline samples of $CuIr_{2-x}Ru_xTe_4$ (0.0 ≤ x ≤ 0.30) were synthesized in two steps by a solid-state reaction method. First, the mixture of high-purity, cleaned fine powders of Cu (99.9 %), Ir (99.9 %), Ru (99.999 %) and Te (99.999 %) in the appropriate stoichiometric ratios were heated in sealed evacuated silica glass tubes at a rate of 1 ºC/min to 850 ºC and held there for 96 hours. Subsequently, the as-prepared powders were reground, re-pelletized, and sintered again, by heating at a rate of 3 ºC/min to 850 ºC and holding there for 72 hours. The identity and phase purity of the samples were determined by powder X-ray diffraction (PXRD) using a Bruker D8 Advance ECO with Cu Kα radiation and a LYNXEYE-XE detector. To determine the unit cell parameters, profile fits were performed on the powder diffraction data in the FULLPROF diffraction suite using Thompson-Cox-Hastings pseudo-Voigt peak shapes.[29] Measurements of the temperature dependent electrical resistivity (4-point method), specific heat, and magnetic susceptibility of the materials were performed in a DynaCool Quantum Design Physical Property Measurement System (PPMS). There was no indication of air-sensitivity of the materials during the study. $T_c$s determined from susceptibility data were estimated conservatively: $T_c$ was taken as the intersection of the extrapolations of the steepest slope of the susceptibility in the superconducting transition region and the normal state susceptibility; for resistivities, the midpoint of the resistivity $ρ(T)$ transitions was taken, and, for the specific heat data, the critical temperatures obtained from the equal area construction method were employed.

**Results and Discussion**

**Fig. 1** show the powder X-ray diffraction patterns at room temperature and fitting unit cell parameters for $CuIr_{2-x}Ru_xTe_4$ (0.0 ≤ x ≤ 0.30). XRD results indicates that the solubility limit for Ru substitution in $CuIr_2Te_4$ is x = 0.30. With higher Ru contents, the cubic $RuTe_2$ phase is obviously found as an impurity. **Fig. 1a** shows the detail refinement results of the selected $CuIr_{1.95}Ru_{0.05}Te_4$ powder. Most of the reflections can be indexed in the $P\bar{3}m1$ space group and the tiny impurity is attributed to the unreacted Ir. The lattice parameters are obtained to be a =3.9360 (3) Å and c = 5.3917 (5) Å. The inset pattern shows that $CuIr_{2-x}Ru_xTe_4$ adopts a disordered trigonal structure, which embodies a two-dimensional (2D) $IrTe_2$ layers and intercalated by Cu between the layers, with Ir partial replacing by Ru. We determined the unit cell parameters by fitting the powder X-ray diffraction data, which were shown at **Fig. 1b**. With the increasing Ru contents, unit cell parameters *a* and *c* decreased linearly. Cell parameters *a* decreased from 3.9397(5) Å ($x = 0$) to 3.9238 (2) Å ($x = 0.30$), meanwhile, parameters *c* decreased from 5.3965 (3) Å ($x = 0$) to 5.3776 (6) Å ($x = 0.30$). The enlargement of (001) peak in **Fig. 1c** shows obvious right shift with the increasing contents of Ru. This phenomenon was also according with the decline of fitting unit cell parameters *c* in **Fig. 1b** by the means of crystal plane spacing formula.

We next perform the temperature dependence of the electrical resistivity $\rho(T)$ and magnetic susceptibility $M(T)$ measurements for $CuIr_{2-x}Ru_xTe_4$ ($0.0 \leq x \leq 0.30$). **Fig. 2a** show the temperature dependence of the normalized electrical resistivities ($\rho/\rho_{300K}$) for the polycrystalline samples of $CuIr_{2-x}Ru_xTe_4$ ($0.0 \leq x \leq 0.30$). At low temperatures (see **Fig. 2b**), a clear, sharp drop of $\rho(T)$ is observed in the $CuIr_{2-x}Ru_xTe_4$ samples ($0.0 \leq x \leq 0.20$) except for the highest doping content sample $CuIr_{1.7}Ru_{0.3}Te_4$, signifying the onset of superconductivity at low temperatures. The transition temperature ($T_c$) slightly rises until it reaches its peak at $x = 0.05$, then, disappears when *x* reaches 0.3. This trend is also clearly seen in the susceptibility data (**Fig. 2c**) - the onset of the negative magnetic susceptibility signaling the systematical superconducting state present a "dome" shape shifts with increasing *x* value for $CuIr_{2-x}Ru_xTe_4$. The superconducting volume fraction can be estimated approximately to be 96 %, which reveals the high purity of the $CuIr_{2-x}Ru_xTe_4$ ($0.0 \leq x \leq 0.20$) samples. In addition, it is obviously seen that there are all no CDW humps for the Ru-doped compounds $CuIr_{2-x}Ru_xTe_4$ ($0.03 \leq x \leq 0.30$) in the temperature-dependent resistivity results, indicating that the CDW state can be suppressed even by small amount substitution Ru for Ir in the host compound $CuIr_2Te_4$, as shown in the **Fig. 2a**. To further prove it, we adopt the measurement of the magnetic susceptibility at applied field of 1T for the smallest doping content compound $CuIr_{1.97}Ru_{0.03}Te_4$. As shown in **Fig. 2d**, unlike the pristine sample $CuIr_2Te_4$, the susceptibility exhibits no change around 250 K, which consistently indicate the CDW transition has been suppressed completely with small amount Ru doping at $x = 0.03$. This significant feature of the interplay between CDW and the superconductivity could be attributed to modifying the Fermi surface and density of states with Ru chemical doping.

Temperature-dependent measurements of the magnetization under incremental magnetic field M(H) were applied to determine the upper critical field $\mu_0H_{c1}(0)$. We choose the optimal doping superconductor to test. **Fig. 3** shows how the $\mu_0H_{c1}(0)$ for the optimal $CuIr_{1.95}Ru_{0.05}Te_4$ compound was determined. First, applied field magnetization measurements M(H) were performed at 1.8, 2.0, 2.2 K and 2.4 K to calculate the demagnetization factor (N). With the hypothesis that the beginning linear response to the magnetic field is perfectly diamagnetic (dM/dH = − 1/4 π) for this bulk superconductor, we obtained the values of demagnetization factor N, of 0.55 – 0.75 (from N = 1/4π$\chi_V$ + 1), where $\chi_V$ = dM/dH is the value of linearly fitted slope for the bottom left corner inset of **Fig. 3**. The experimental data can be fitted with the formula $M_{fit} = a + bH$ at low magnetic fields, where a is an intercept and b is a slope from fitting the low magnetic field magnetization measurements data. The up-right corner inset of **Fig. 3** shows the M(H) − $M_{fit}$ data versus the magnetic field(H). $\mu_0H_{c1}^*$ was determined at the field when M deviates by ∼ 1 % above the fitted data ($M_{fit}$), as is the common practice.[30] We can calculate the lower critical field $\mu_0H_{c1}(T)$ in the consideration of the demagnetization factor (N), *via* using the relation $\mu_0H_{c1}(T) = \mu_0H_{c1}^*(T)/ (1 − N)$. [31-33] The main panel of **Fig. 3** reveals the $\mu_0H_{c1}(T)$ as the function of temperature for $CuIr_{1.95}Ru_{0.05}Te_4$. We estimated the $\mu_0H_{c1}(0)$ by fitting the $\mu_0H_{c1}(T)$ data *via* the formula $\mu_0H_{c1}(T) = \mu_0H_{c1}(0) [1 − (T/T_c)^2]$, which was shown by the black solid lines. The obtained zero-temperature lower critical field $\mu_0H_{c1}(0)$ for $CuIr_{1.95}Ru_{0.05}Te_4$ was 0.098 T (**Table 1**), which is higher than that of the host compound $CuIr_2Te_4$.

With the purpose of estimating the critical field $\mu_0H_{c2}(0)$, we examined temperature dependent electrical resistivity at various applied fields $\rho(T, H)$ for $CuIr_{1.95}Ru_{0.05}Te_4$ sample. **Fig. 4** exhibits the $\rho(T, H)$ measurement data for $CuIr_{1.95}Ru_{0.05}Te_4$. **Inset of Fig. 4** shows upper critical field values $\mu_0H_{c2}$ plotted *vs* temperature with $T_c$s obtained from resistivity at different applied fields. The $\mu_0H_{c2}$ vs $T$ curve near $T_c$ of CuIr$_{1.95}$Ru$_{0.05}$Te$_4$ sample shows the well linearly fitting, which is represented by solid line. The value of fitting data slope (d$H_{c2}$/d$T$) of CuIr$_{1.95}$Ru$_{0.05}$Te$_4$ sample was shown in **Table 1**. We can estimate the zero-temperature upper critical field ($\mu_0H_{c2}$(T)) of 0.247 T for CuIr$_{1.95}$Ru$_{0.05}$Te$_4$ from the data, using the Werthamer-Helfand-Hohenberg (WHH) expression formula $\mu_0H_{c2}$(T) = -0.693$T_c$ (d$H_{c2}$/d$T_c$) for the dirty limit superconductivity.[33-37] The obtained $\mu_0H_{c2}$(T) for CuIr$_{1.95}$Ru$_{0.05}$Te$_4$ is two times higher than that of the pristine CuIr$_2$Te$_4$, as summarized in **Table 1**. In addition, the Pauli limiting field ($\mu_0H^P$(T)) of CuIr$_{1.95}$Ru$_{0.05}$Te$_4$ can be calculated from $\mu_0H^P$(T) = 1.86$T_c$. The calculated values of $\mu_0H^P$(T) was also larger than that of the host compound CuIr$_2$Te$_4$. Then, with this formula $\mu_0H_{c2}(T) = \frac{\phi_0}{2\pi\xi_{GL}^2}$, where $\phi_0$ is the flux quantum, the Ginzburg-Laudau coherence length ($\xi_{GL}(0)$) was calculated ~ 36.3 nm for CuIr$_{1.95}$Ru$_{0.05}$Te$_4$ (**Table 1**).

Temperature-dependent specific-heat measurements were performed with the exception of magnetic susceptibility and resistivity measurements to confirm that superconductivity is an intrinsic property of CuIr$_{1.95}$Ru$_{0.05}$Te$_4$. **Fig. 5** (main panel) plots $C_p/T$ vs $T^2$ in zero and 3 tesla applied field in temperature range of 2 - 10 K. The relationship for the $C_p/T$ vs T in zero applied field near the transition temperature was further plotted in the inset of **Fig 5**. As it can be seen, there is large anomaly hump in the specific heat data, which agrees with bulk superconductivity in CuIr$_{1.95}$Ru$_{0.05}$Te$_4$. The superconducting transition temperature ($T_c$) can be confirmed by equal-entropy constructions of the idealized specific-heat capacity jump (shown with purple shading). The $T_c$ of CuIr$_{1.95}$Ru$_{0.05}$Te$_4$ was determined to be 2.72 K, which is very close to the $T_c$s obtained from the resistivity and magnetic susceptibility measurements. Further, we got the values of $\gamma$ and $\beta$ (**Fig. 5**) from fitting the data got under 3 tesla applied field in temperature range of 2 - 10 K. The normalized specific heat jump value $\Delta C/\gamma T_c$ obtained from the data (**inset of Fig. 5**) was 1.51 for CuIr$_{1.95}$Ru$_{0.05}$Te$_4$, which is higher than the Bardeen-Cooper-Schrieffer (BCS) weak-coupling limit value (1.43), confirming bulk superconductivity. Then we obtain the Debye temperature by the formula $\Theta_D$ =

$(12\pi^4 nR/5\beta)^{1/3}$ by using the fitted value of $\beta$, where $n$ is the number of atoms per formula unit and $R$ is the gas constant. Thus, we can estimate the electron-phonon coupling constant ($\lambda_{ep}$) by using the Debye temperature ($\Theta_D$) and critical temperature $T_c$ from the inverted McMillan formula: $\lambda_{ep} = \frac{1.04 + \mu^* \ln\left(\frac{\Theta_D}{1.45 T_c}\right)}{(1-1.62\mu^*)\ln\left(\frac{\Theta_D}{1.45 T_c}\right) - 1.04}$ [33]. This resultant $\lambda_{ep}$ is 0.67, suggesting that CuIr$_{1.95}$Ru$_{0.05}$Te$_4$ belongs to a strongly electron-phonon coupled superconductor. The electron density of states at the Fermi level ($N(E_F)$) can be calculated from $N(E_F) = \frac{3}{\pi^2 k_B^2 (1+\lambda_{ep})}\gamma$ with the $\gamma$ and $\lambda_{ep}$. We got the value that $N(E_F)$ = 2.92 states/eV f.u. for CuIr$_{1.95}$Ru$_{0.05}$Te$_4$ and $N(E_F)$ = 2.72 states/eV f.u. for CuIr$_2$Te$_4$ (**Table 1**). This result indicates that the higher density of electronic states at the Fermi energy matched the higher transition temperature due to the Ru doping into CuIr$_2$Te$_4$.

To further understand the effect of doping on superconducting transition temperature, we have established the electronic phase diagram plotted $T_c$s vs $x$ doping content for CuIr$_{2-x}$Ru$_x$Te$_4$ ($0.0 \leq x \leq 0.30$), as shown in **Fig. 6**. All the $T_c$s were obtained from the temperature dependence of the normalized ($\rho/\rho_{300K}$) resistivities and magnetic susceptibility data. From the phase diagram we can easily find that the $T_c$ vs. $x$ content present a "dome-like" shape. Using Ru chemical doping as finely controlled tuning parameters, the CDW state has been mediately surprised, meanwhile superconducting transition temperature ($T_c$) rises to the first peak ($x$ = 0.05) and then decreases until it reaches its minimum value at $x$ = 0.3. Nevertheless, the season why the CDW state can be suppressed by Ru doping so quickly has not yet been studied. Through systematic research the doping system of CuIr$_2$Te$_4$, we found that materials' electronic structure can by effected by the doping content consequently affect their physical properties like superconducting transition temperature, also there is large room for further exploration the interplay of CDW and superconductivity in $AB_2X_4$ system.

## Conclusion

Here the solid solutions CuIr$_{2-x}$Ru$_x$Te$_4$ ($0.0 \leq x \leq 0.3$) have been successfully synthesized *via* solid-state reaction to study the effect of the *B*-site substitution on the superconductivity. The structural and superconductivity properties for this system was evaluated systematically by means of powder x-ray diffraction (XRD), magnetization, resistivity and specific-heat measurements. XRD analysis reveals that $CuIr_{2-x}Ru_xTe_4$ (0.0 ≤ $x$ ≤ 0.3) crystallized a disordered trigonal structure with space group $P\bar{3}m1$ (No. 164). Specific-heat, isothermal magnetization {M(H)} and magneto-transport {ρ(T, H)} measurements results signify that the optimal doping content compound $CuIr_{1.95}Ru_{0.05}Te_4$ is a strongly electron-phonon coupled type-II superconductor with $T_c$ ≈ 2.79(1) K, a lower critical field $H_{c1}(0)$ = 980 Oe and an upper critical field, $H_{c2}(0)$ = 2470 Oe. Finally, we have established a "dome-like" shape electronic phase diagram, in which CDW-superconducting transition temperature as a function of Ru doping content $x$. We can easily find that the CDW has been suppressed immediately at $x$ = 0.03 and the superconducting transition temperature ($T_c$) rises to the first peak ($x$ = 0.05) and then decreases until it reaches its minimum value at $x$ = 0.3, which displays a good material platform for further study the competition between CDW and superconductivity.

**Acknowledgment**

The authors thank T. Klimczuk, B. Shen and T. R. Chang for valuable discussions. H. X. Luo acknowledges the financial support by Natural Science Foundation of China (No. 21701197 and No. 11922415). D. X. Yao are supported by NKRDPC Grants No. 2017YFA0206203, No. 2018YFA0306001, NSFC-11574404, and Leading Talent Program of Guangdong Special Projects.

**Reference:**
[1] T. J. Coutts, D. L. Young, X. Li, W. P. Mulligan, and X. Wu, J. Vac. Sci. Technol. A**18**, 2646 (2000).
[2] M. Dekkers, G. Rijnders, and D. H. A. Blank, Appl. Phys. Lett. **90**, 021903 (2007).
[3] C. A. Hoel, T. O. Mason, J.-F. Gaillard, and K. R. Poeppelmeier, Chem. Mater. **22**, 3569 (2010).
[4] F. K. Lotgering, Philips Res. Rep., **11**, 190 (1956).
[5] S. Nagata, T. Hagino, Y. Seki, and T. Bitoh, Phys. B, **194-196**, 1077 (1994).

[6] P. G. Radaelli, Y. Horibe, M. J. Gutmann, H. Ishibashi, C. H. Chen, R. M. Ibberson, Y. Koyama, Y. S. Hor, V. Kiryukhin, and S. W. Cheong, Nature, **416**, 155 (2002).

[7] T. Hagino, Y. Seki, N. Wada, S. Tsuji, T. Shirane, K. I. Kumagai, and S. Nagata, Phys. Rev. B, **51**, 12673 (1995).

[8] M. Ito, A. Taira, and K. Sonoda, Acta Physica Polonica A., **131**, 6 (2017).

[9] T. Hagino, Y. Seki, and S. Nagata, Phys. C, **235–240**, 1303(1994).

[10] T. Furubayashi, T. Matsumoto, T. Hagino, and S. Nagata, J. Phys. Soc. Jpn., **63**, 3333 (1994).

[11] T. Hagino, T. Tojo, T. Atake, and S. Nagata, Philos. Mag. **B71**, 881 (1995).

[12] T. Oda, M. Shirai, N. Suzuki, and K. Motizuki, J. Phys.: Condens. Matter, **7**, 4433 (1995).

[13] G. Oomi, T. Kagayama, I. Yoshida, T. Hagino, and S. Nagata, J. Magn. Magn. Mater., **140–144**, 157 (1995).

[14] T. Furubayashi, T. Kosaka, J. Tang, T. Matsumoto, Y. Kato, and S. Nagata, J. Phys. Soc. Jpn., **66** 1563(1997).

[15] M. Robbins, R. H. Willens, and R. C. Miller, Solid State Commun., **5**, 933 (1967).

[16] F. K. Lotgering, and R. P. Van Stapele, J. Appl. Phys., 39, 417 (1968).

[17] F. K. Lotgering, J. Phys. Chem. Solids, **30**, 1429 (1969).

[18] F. J. DiSalvo and J. V. Waszczak, Phys. Rev. B, **26**, 2501 (1982).

[19] R. N. Shelton, D. C. Johnston, and H. Adrian, Solid State Commun., **20**, 1077 (1976).

[20] R. M. Fleming, F. J. DiSalvo, R. J. Cava and J. V. Waszczak, Phys. Rev. B, **24**, 2850 (1981).

[21] N. L. Nagard, A. Katty, G. Collin, O. Gorochov, and A. Willig, J. Solid State Chem., **27**, 267 (1979).

[22] S. V. Smaalen, Acta Crystallogr. **A61**, 51 (2005).

[23] S. Nagata, N. Kijima, S. Ikeda, N. Matsumoto, R. Endoh, S. Chikazawa, I. Shimono, and H. Nishihara, J. Phys. Chem. Solids, **60**, 163 (1999).

[24] D. Yan, Y. J. Zeng, G. H. Wang, Y. Y. Liu, J. J. Yin, T.-R. Chang, H. Lin, M. Wang, J. Ma, S. Jia, D.-X. Yao, H. X. Luo, arXiv:1908.05438 (2019).

[25] S. Nagata, N. Matsumoto, Y. Kato, T. Furubayashi, T. Matsumoto, J. P. Sanchez, and P. Vulliet, Phys. Rev. B, **58**, 6844 (1998).

[26] N. Matsumoto, R. Endoh, S. Nagata, T. Furubayashi, and T. Matsumoto, Phys. Rev. B, **60**, 5258 (1999).

[27] H. Suzuki, T. Furubayashi, G. Cao, H. Kitazawa, A. Kamimura, K. Hirata and T. Matsumoto, J. Phys. Soc. Jpn., **68**, 2495 (1999).

[28] H. X. Luo, T. Klimczuk, L. Müchler, L. Schoop, D. Hirai, M. K. Fuccillo, C. Felser, and R. J. Cava, Phys. Rev. B, **87**, 214510 (2013).

[29] J. Rodríguez-Carvajal, Comm. Powder Diffr., **26**, 12 (2001).

[30] H. X. Luo, W. W. Xie, J. Tao, I. Pletikosic, T. G. Valla, S. Sahasrabudhe, G. Osterhoudt, E. Sutton, K. S. Burch, E. M. Seibel, J. W. Krizan, Y. M. Zhu, and R. J. Cava, Chem. Mater., **28**, 1927 (2016).

[31] M. J. Winiarski, B. Wiendlocha, S. Gołąb, S. K. Kushwaha, P. Wiśniewsk, D. Kaczorowski, J. D. Thompson, R. J. Cava, and T. Klimczuk, Phys. Chem. Chem. Phys.,

**18**, 21737 (2016).

[32] C. S. Yadav, and P. L. Paulose, New J. Phys., **11**, 103046 (2009).

[33] W. L. McMillan, Phys. Rev., **167**, 331(1968).

[34] J. A. Wilson, A. S. Barker, F. J. Di Salvo, and J. A. Ditzenberger, Phys. Rev. B, **18**, 2866 (1978).

[35] W. Kohn, Phys. Rev. Lett., **19**, 439 (1967).

[36] N. R. Werthamer, E. Helfand, and P. C. Hohenberg, Phys. Rev., **147**, 415 (1967).

[37] V. Z. Kresin, and S. A. Wolf, Plenum Press, New York and London, 150 (1990).

**Table 1. Comparison of superconducting parameters in $AB_2X_4$ superconductors**

| Material | CuIr$_{1.95}$Ru$_{0.05}$Te$_4$ | CuIr$_2$Te$_4$ | CuRh$_2$S$_4$ | CuRh$_2$Se$_4$ | Cu$_{0.7}$Zn$_{0.3}$Ir$_2$S$_4$ | CuIr$_{1.6}$Pt$_{0.4}$Se$_4$ |
|---|---|---|---|---|---|---|
| $T_c$ (K) | 2.79 | 2.50 | 4.7 | 3.5 | 3.4 | 1.76 |
| $\gamma$ (mJ mol$^{-1}$ K$^{-2}$) | 11.52 | 10.57 | 26.9 | 21.4 | | 16.5 |
| $\beta$ (mJ mol$^{-1}$ K$^{-4}$) | 2.54 | 2.15 | | | | 1.41 |
| $\Theta_D$ (K) | 174.8(1) | 185.5 (2) | 258 | 218 | | 212 |
| $\Delta C/\gamma T_c$ | 1.51 | 1.82 | 1.89 | 1.68 | | 1.58 |
| $\lambda_{ep}$ | 0.67 | 0.65 | 0.66 | 0.63 | | 0.57 |
| $N(E_F)$ (states/eV f.u) | 2.92 | 2.72 | | | | 4.45 |
| $-dH_{c2}/dT$ (T/K) | 0.125 | 0.066 | 0.614 | 0.181 | | 2.62 |
| $\mu_0 H_{c2}$(T) | 0.247 | 0.12 | 2.0 | 0.44 | | 3.2 |
| $\mu_0 H^P$(T) | 5.24 | 4.65 | 8.74 | 6.51 | 6.32 | 3.27 |
| $\mu_0 H_{c1}$(T) | 0.098 | 0.028 | | | | |
| $\xi_{GL}(0)$ (nm) | 36.3 | 52.8 | | | - | 0.96 |

**Table 2.** Rietveld refinement structural parameters of CuIr$_{1.95}$Ru$_{0.05}$Te$_4$. Space group $P\bar{3}m1$ (No. 164), $a$ = b =3.9360(1) Å and $c$ = 5.3917(2) Å, $R_p$ = 6.29 %, and $R_{wp}$ = 9.90 %.

| Label | x | y | z | Site | OCC. |
|---|---|---|---|---|---|
| Ir | 0.00000 | 0.00000 | 0.00000 | 1a | 0.950 |
| Ru | 0.00000 | 0.00000 | 0.00000 | 1a | 0.050 |
| Te | 0.33330 | 0.66670 | 0.2308(4) | 2d | 1.000 |
| Cu | 0.00000 | 0.00000 | 0.50000 | 1b | 0.500 |

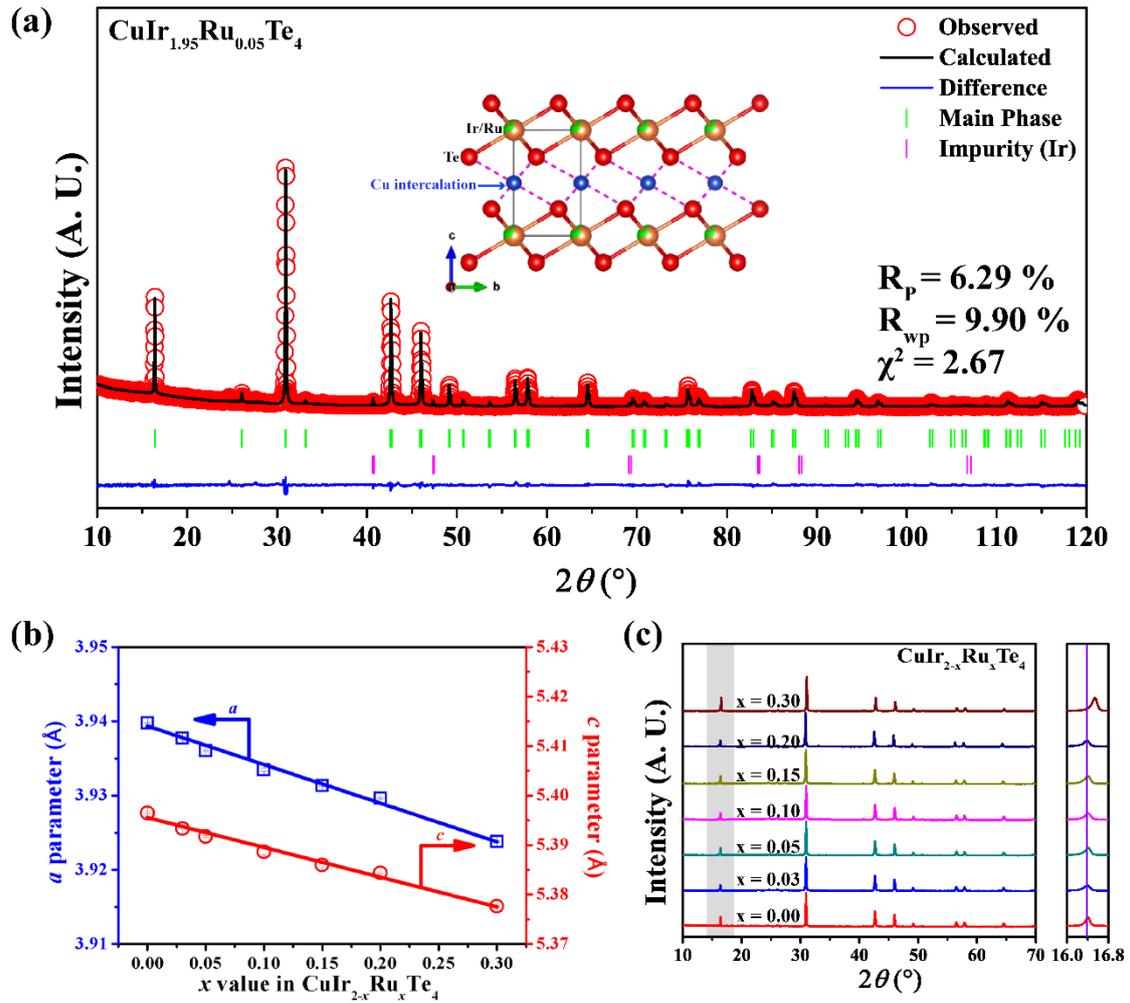

**Figure 1. Structural and chemical characterization of CuIr$_{2-x}$Ru$_x$Te$_4$.** (A) Powder XRD patterns (Cu Kα) for the CuIr$_{2-x}$Ru$_x$Te$_4$ samples studied (0.0 ≤ $x$ ≤ 0.30). Inset shows the enlargement of peak (001). (B) The evolution of lattice parameter $a$ and $c$ of CuIr$_{2-x}$Ru$_x$Te$_4$. (C) Powder XRD pattern with Rietveld refinement for CuIr$_{1.95}$Ru$_{0.05}$Te$_4$.

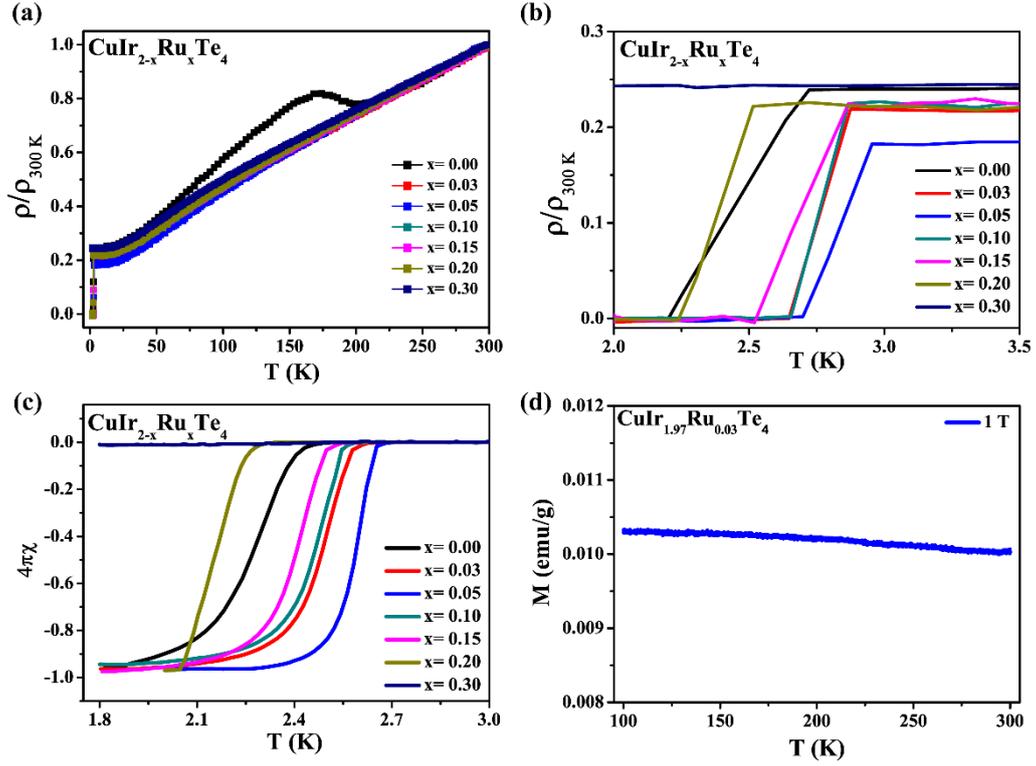

**Figure 2. Transport characterization of the normal states and superconducting transitions for CuIr$_{2-x}$Ru$_x$Te$_4$.** (a) The temperature dependence of the resistivity ratio ($\rho/\rho_{300K}$) for polycrystalline CuIr$_{2-x}$Ru$_x$Te$_4$ ($0.0 \leq x \leq 0.30$). (b) The temperature dependence of the resistivity ratio ($\rho/\rho_{300K}$) for polycrystalline CuIr$_{2-x}$Ru$_x$Te$_4$ at low temperature. (c) Magnetic susceptibilities for CuIr$_{2-x}$Ru$_x$Te$_4$ ($0.0 \leq x \leq 0.30$) at the superconducting transitions; applied DC fields are 20 Oe. (d) Magnetic susceptibility of CuIr$_{1.97}$Ru$_{0.03}$Te$_4$ as a function of temperature at applied field of 1 tesla.

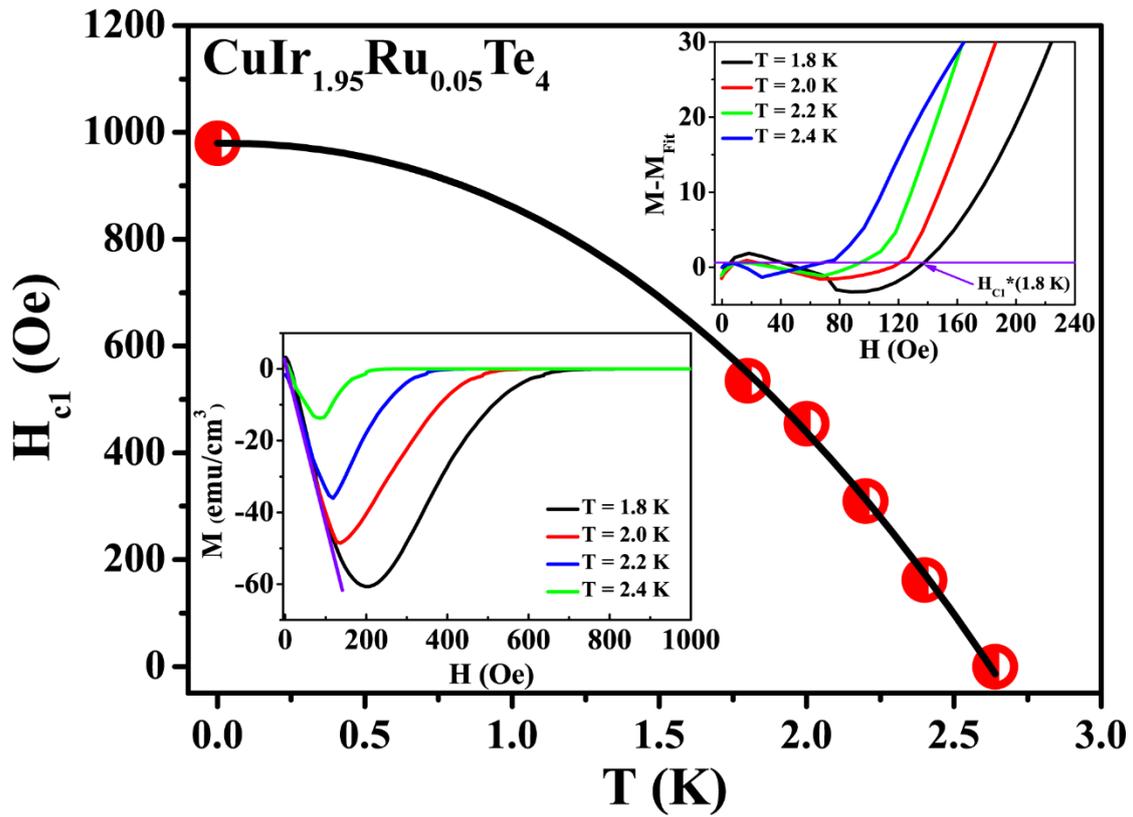

**Figure 3**. Temperature dependence of the lower critical field ($\mu_0 H_{c1}$) for $CuIr_2Te_4$. Bottom left corner inset shows magnetic susceptibility at low applied magnetics fields at various temperatures for $CuIr_{1.95}Ru_{0.05}Te_4$. Up right inset shows M-M$_{fit}$ vs H.

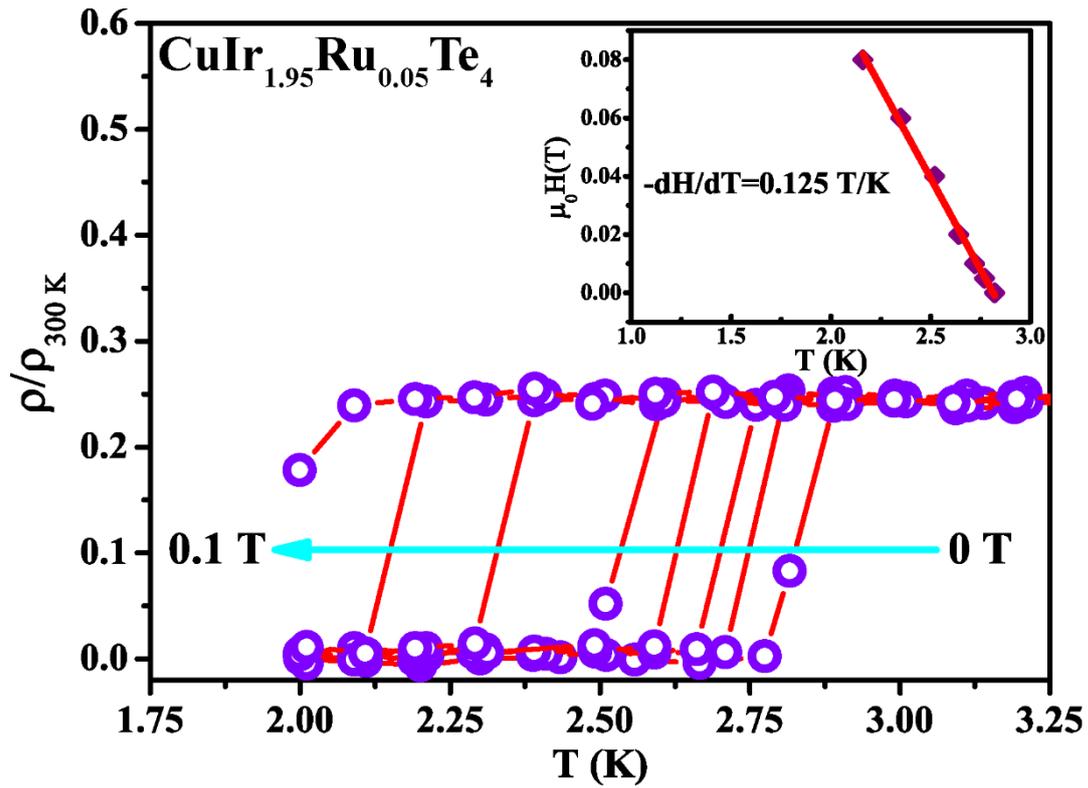

**Figure 4.** Low temperature resistivity at various applied fields for CuIr$_{1.95}$Ru$_{0.05}$Te$_4$. Inset shows $\mu_0H(T)$ at different $T_c$s, red solid line shows linearly fitting to the data to estimate $\mu_0H_{c2}$.

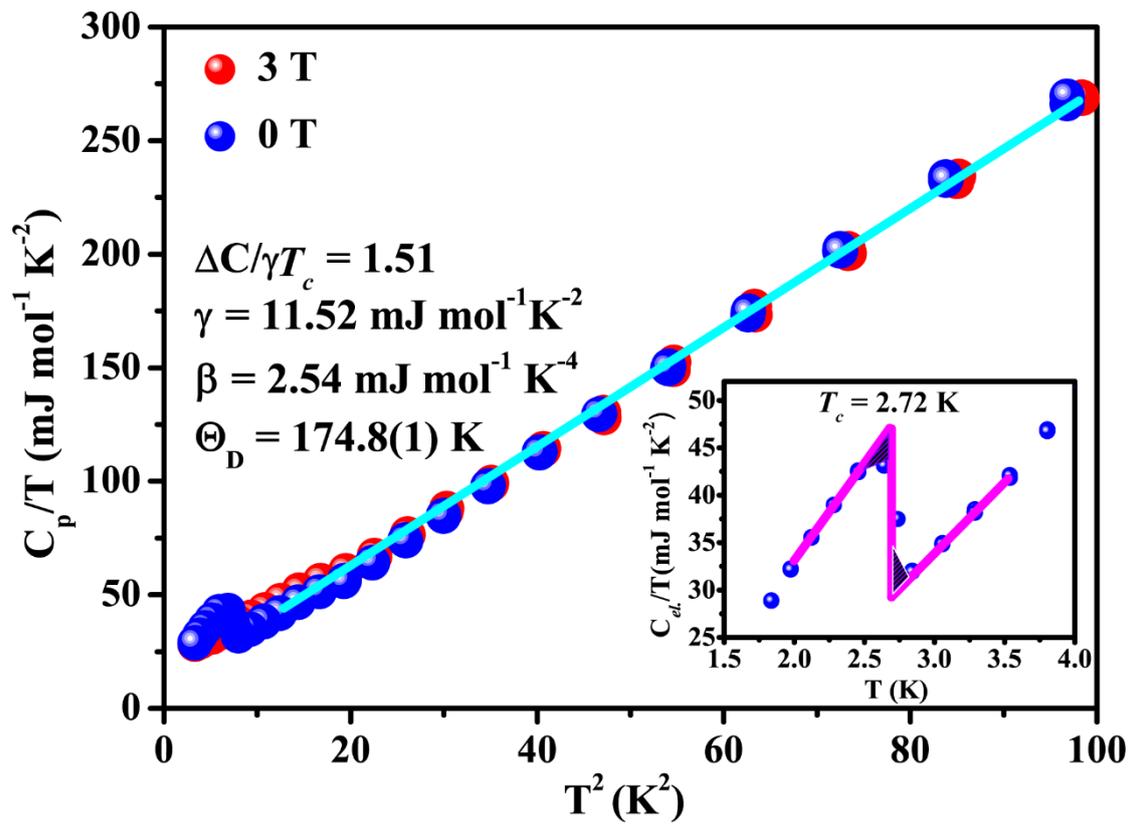

**Figure 5. Heat capacity characterization of CuIr$_{1.95}$Ru$_{0.05}$Te$_4$.** Debye temperature of CuIr$_{1.95}$Ru$_{0.05}$Te$_4$ obtained from fits to data in applied field. Inset shows the heat capacity through the superconducting transition without applied magnetic field for CuIr$_{1.95}$Ru$_{0.05}$Te$_4$.

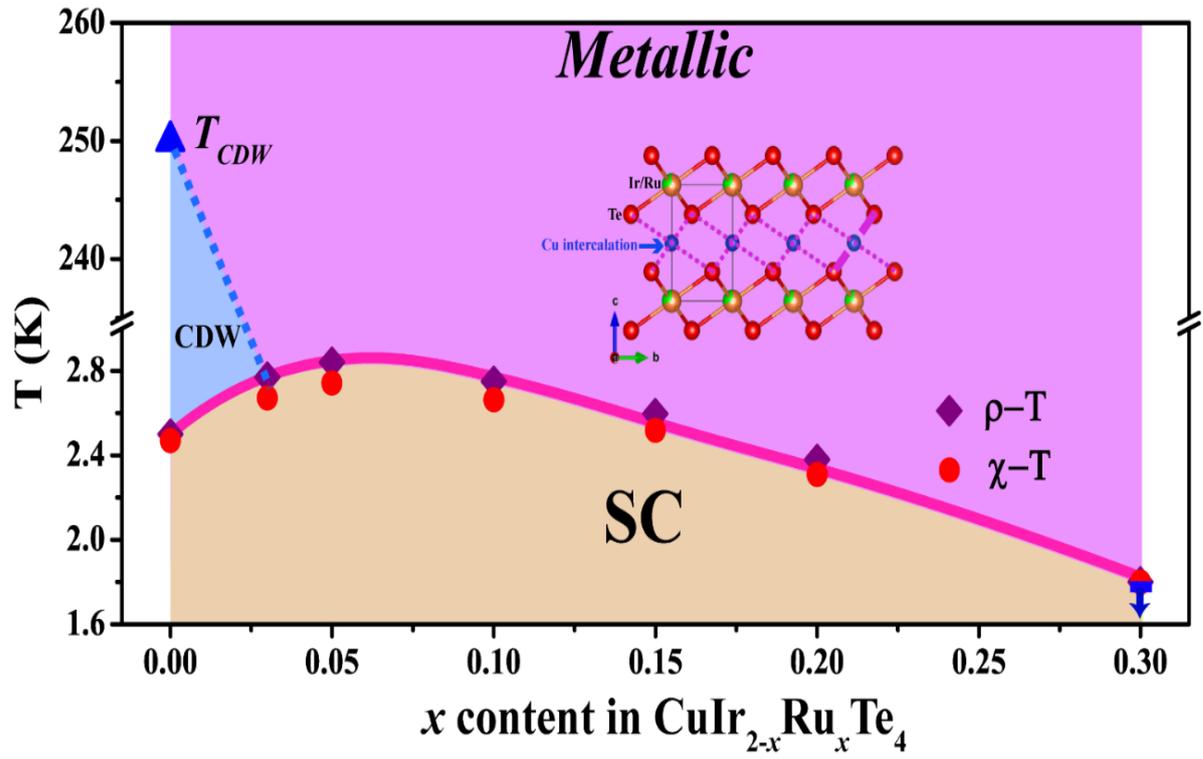

**Figure 6.** The electronic phase diagram for CuIr$_{2-x}$Ru$_x$Te$_4$ (0.0 ≤ $x$ ≤ 0.30).